\documentclass[a4paper,12pt]{article}
\usepackage{amsfonts,amsthm,amsmath,amssymb,graphicx}
\usepackage{braket}
\usepackage{bm}
\usepackage{float}
\usepackage{mathrsfs}
\usepackage{booktabs}
\usepackage{makeidx}
\usepackage{color}
\usepackage{hangcaption}
\usepackage{atbegshi}
\usepackage{comment}
\usepackage[dvipdfmx,
bookmarks=true,bookmarksnumbered=true,%
linkcolor=blue,anchorcolor=blue,urlcolor=blue,
]{hyperref}

\begin{document}

\newtheorem{definition}{Definition}[section]
\newcommand{\be}{\begin{equation}}
\newcommand{\ee}{\end{equation}}
\newcommand{\bea}{\begin{eqnarray}}
\newcommand{\eea}{\end{eqnarray}}
\newcommand{\LE}{\left[}
\newcommand{\R}{\right]}
\newcommand{\nn}{\nonumber}
\newcommand{\Tr}{\text{Tr}}
\newcommand{\N}{\mathcal{N}}
\newcommand{\G}{\Gamma}
\newcommand{\vf}{\varphi}
\newcommand{\LL}{\mathcal{L}}
\newcommand{\HH}{\mathcal{H}}
\newcommand{\arctanh}{\text{arctanh}}
\newcommand{\up}{\uparrow}
\newcommand{\down}{\downarrow}
\newcommand{\ketbra}[1]{\left|#1\right>\left<#1\right|}
\newcommand{\rd}{\partial}
\newcommand{\de}{\partial}
\newcommand{\ba}{\begin{eqnarray}}
\newcommand{\ea}{\end{eqnarray}}
\newcommand{\db}{\bar{\partial}}
\newcommand{\we}{\wedge}
\newcommand{\ca}{\mathcal}
\newcommand{\lr}{\leftrightarrow}
\newcommand{\f}{\frac}
\newcommand{\s}{\sqrt}
\newcommand{\vp}{\varphi}
\newcommand{\hvp}{\hat{\varphi}}
\newcommand{\tvp}{\tilde{\varphi}}
\newcommand{\tp}{\tilde{\phi}}
\newcommand{\ti}{\tilde}
\newcommand{\ap}{\alpha}
\newcommand{\pr}{\propto}
\newcommand{\mb}{\mathbf}
\newcommand{\ddd}{\cdot\cdot\cdot}
\newcommand{\no}{\nonumber \\}
\newcommand{\la}{\langle}
\newcommand{\lb}{\rangle}
\newcommand{\ep}{\epsilon}
\newcommand{\al}{\alpha'}
\newcommand{\ddbp}{\mbox{D}p-\overline{\mbox{D}p}}
\newcommand{\ddbt}{\mbox{D}2-\overline{\mbox{D}2}}
\newcommand{\ov}{\overline}
 \def\prt{\partial}
 \def\vep{\varepsilon}
 \def\tr{\text{tr}}
 \def\we{\wedge}
 \def\lr{\leftrightarrow}
 \def\f {\frac}
 \def\ti{\tilde}
 \def\ap{\alpha}
 \def\pr{\propto}
 \def\mb{\mathbf}
 \def\ddd{\cdot\cdot\cdot}
 \def\no{\nonumber \\}
 \def\la{\langle}
 \def\lb{\rangle}
 \def\ep{\epsilon}

\def\RR{{\bf R}}
\def\FF{{\cal F}}
\renewcommand{\*}{ &=& }
\newcommand{\xsim}{ &\sim& }

\newcommand{\sfrac}[2]{{{#1}/{#2}}}

\makeatletter
    \renewcommand{\theequation}{%
    \thesection.\arabic{equation}}
    \@addtoreset{equation}{section}
  \makeatother

\begin{titlepage}
\thispagestyle{empty}

\begin{flushright}
YITP-14-105\\
IPMU14-0358\\
\end{flushright}

\vspace{.4cm}
\begin{center}
\noindent{\Large \textbf{Notes on Entanglement Entropy in String Theory}}\\
\vspace{2cm}

Song He $^{a,b}$,
Tokiro Numasawa $^{a}$,
Tadashi Takayanagi $^{a,c}$,
Kento Watanabe $^a$
\vspace{1cm}

{\it
 $^{a}$Yukawa Institute for Theoretical Physics (YITP),\\
Kyoto University, Kyoto 606-8502, Japan\\
$^{b}$State Key Laboratory of Theoretical Physics, Institute of Theoretical Physics, \\ Chinese Academy of Science, Beijing 100190, P. R. China\\
$^{c}$Kavli Institute for the Physics and Mathematics of the Universe (Kavli IPMU),\\
University of Tokyo, Kashiwa, Chiba 277-8582, Japan\\
}

\vskip 2em
\end{center}

\vspace{.5cm}
\begin{abstract}
In this paper, we study the entanglement entropy in string theory in the simplest setup 
of dividing the nine dimensional space into two halves. This corresponds to the leading quantum correction to the horizon entropy in string theory on the Rindler space. This entropy is also called the conical entropy and includes surface term contributions. We first derive a new simple formula of the conical entropy for any free higher spin fields. Then we apply this formula to computations of conical entropy in open and closed superstring. In our analysis of closed string, we study the twisted conical entropy defined by making use of string theory on Melvin backgrounds. This quantity is easier to calculate owing to the folding trick. Our analysis shows that the entanglement entropy in closed superstring is UV finite owing to the string scale cutoff.

\end{abstract}

\end{titlepage}

\section{Introduction}
Entanglement entropy (EE) is a useful measure of the degrees of freedom in quantum many-body systems. For example,
we can use EE to detect the central charges \cite{HLW,CalCa,RTa,Sold,MySi,CHM}. Moreover, we can detect the topological degrees of freedom of the topological field theories \cite{KiPre,LeWe}.
In recent studies, it has also been revealed that the EE can measure the degrees of freedom of local operators \cite{NNT,SNTW,CNT,Nozaki,ABGH,CSST}.

To calculate EE in field theories, there is a well-known method called the replica method, by which the calculation of EE returns to the calculation of partition functions on the space with conical defects. In field theories, entanglement entropy has the area law UV divergences \cite{Sredniki,BKLS} because each Hilbert space exists on the point in a space and there are contributions from entanglement between infinitely small separated Hilbert spaces.
Therefore usually we introduce the UV cut off parameter $\vep$ (lattice spacing) and we only consider the entanglement between the Hilbert space more separated than $\vep$.
 In the string theory, the theory has a parameter of sting length $l_s=\s{\alpha'}$ beyond which
there is no more small structure. Thus it is a natural question how the results are changed if we consider the string theory instead of such local quantum field theories. It is natural to expect that the cut off scale $\vep$ is replaced by string parameter $\alpha '$ and that the EE in string theory becomes UV finite. By combining this observation with the area law, we may expect the following behavior of the EE in superstring (we assume $V_8\gg \al^4$):
\be
S_{A}=s\cdot\f{V_{8}}{\al^{4}}+\ddd,  \label{areastring}
\ee
where $V_8$ is the area of the boundary of the subsystem $A$ and $s$ is a $O(1)$ constant; 
the omitted terms $\ddd$ denote the sub-leading contributions in the limit $\al\to 0$. Indeed this interesting problem has been discussed in early papers \cite{Sus,Suss,Suk,SU,Dah1,Dah2,LS,Emp} because the entanglement entropy in string theory is expected to be equal to the quantum corrections to the black hole entropy. Nevertheless, there have been no explicit evaluations of EE in string theory done until now as far as the authors of the present paper know. This is the main motivation of this paper.

In this paper we focus on the EE between a half of the total space and the other half in the ten dimensional flat space $R^{1,9}$. At first, one may worry that it may be difficult to define EE in string theory using this real space division because the theory itself is non-local. However, there is one clear definition by the replica method, generalizing computations of EE in field theories, and we will employ this idea. In this paper, we will not get into serious considerations of the original definition of EE in terms of Hilbert space factorizations in string theory. In the replica method, we need partition functions on $n$-replicated manifolds.
If we replace the replica number $n$ with a fractional number $1/N$, then the manifold behaves like
an orbifold $C/\mathbb{Z}_N$ times $R^8$. Though we are not familiar with string theory on the spacetime defined by the $n$-sheeted replica manifold, we know string theory on orbifold backgrounds very well.
Therefore we would like to evaluate the EE in string theory
by analyzing orbifolds. Because the orbifolds break spacetime supersymmetries, tachyonic modes appear from the twisted sectors \cite{Dah1,LS} and thus the R\'enyi entanglement entropy (REE) naively gets divergent. However, to calculate (von-Neumann) entanglement entropy, we only need the derivative w.r.t $N$
at $N=1$. Since we only need the behavior around the $N=1$ point where the supersymmetry is kept,
we can still expect that in superstring, the entanglement entropy becomes finite. To properly analyze the R\'enyi entanglement entropy we would need to take into account the effect of closed string tachyon condensation properly (see e.g. \cite{APS,Da,HKMM,DGHM,HMT} for earlier analysis of localized closed string tachyon condensations ).

Since string theory contains infinitely many higher spin fields, the EE in string theory contains contributions from all such higher spin fields. In the calculations of EE for spin one (Maxwell field) or higher spin fields, there is a subtle
issue related to the surface term contributions. For example, in the Maxwell theory, if we calculation the entanglement entropy using the replica method, we find that the resulting entropy is negative due to a surface term \cite{Kab}. Recently this term is interpreted as that produced by edge modes localized at the entangling surface \cite{DWa,Hua}. A similar surface term exists for spin $3/2$ and $2$ fields and moreover there appear new topological contributions to the EE for spin $2$ fields as shown in \cite{FM}.
Such a entropy, which is directly obtained from the replica method and can be negative, is called the conical entropy. This conical entropy evaluated at one loop (torus) string amplitudes gives the quantum corrections to black hole entropy, while the Bekenstein-Hawking formula \cite{Bek,Haw} is explained by the conical entropy at tree level (sphere) string amplitudes \cite{SU,Solodukhin}.
Since the conical entropy (at one loop) has negative surface contributions,
it has been conjectured in \cite{SU} that the conical entropy may vanish in string theory backgrounds with enough supersymmetries i.e. $s=0$ in (\ref{areastring}). More generally, the conical entropy is identified with quantum corrections to the holographic entanglement entropy formula \cite{RTa,RT,HRT} as recently argued in \cite{FLM}.
In this paper, we will calculate the conical entropy at one loop level for any free higher spin fields using our orbifold method, which largely generalize the result \cite{FM}, and then apply our results to conical entropies in open and closed (super)string theory.

To calculate the conical entropy in closed string theory, we need to evaluate torus partition functions as the explicit function of $N$ in order to obtain an analytical continuation.
In closed string theory, there are not only
the $\mathbb{Z}_N$ projection but also the summation over twisted sectors. Thus we need to
perform two summations in order to analytically continue w.r.t $N$, which is technically
very difficult. To avoid the double summation about $g \in \mathbb{Z}_N$, we compactify one direction to $S^1$ and also twist this direction by $\mathbb{Z}_N$.
Then, type II string theory on this orbifold is related to the string theory
in the Melvin background \cite{DGGH1,DGGH2,Tsy,Bosonic,RuTs,TaUe,brane,DGHM} under the T-duality transformation.
Indeed, it was shown that the small radius limit of string theory on the Melvin background is
equivalent to the $C/\mathbb{Z}_N$ orbifold theory \cite{TaUe,brane}.
This definition of conical entropy leads to a modification of the standard conical entropy because of the twisting in the $S^1$ direction. Therefore we call this new quantity the twisted conical entropy.
In this paper we will analyze the (twisted) conical entropy in type II string theory.

The plan of this paper is as follows. In section 2 we will compute the conical entropy for free field theories with arbitrarily high spins. In section 3 we will investigate the conical entropy for open string theories in bosonic string and superstring. In section 4, we will define the conical entropy for closed superstring theory. We will explicitly evaluate the twisted conical entropy in type II superstring theory using Melvin background. We will analyze its convergence. We will also comment about the effect of twisting about $S^1$ direction. In section 5 we draw conclusions and discuss the future problems. In appendix, we follow the technical detail about the folding trick and about the summation
needed in section 4.

\section{EE for free fields with arbitrary higher spins}

Before we start to analyze entanglement entropy (EE) in string theory, we would like to study the EE in free field theories with arbitrary spins using the first quantization approach. The EE in string theory can be analyzed in the parallel way as we will see later sections.

We express the partition function of a free field theory on the spacetime $M$ at the first quantized level as $Z^f(M)$.
In the second quantization approach, this corresponds to the logarithm of partition function $Z^s(M)$ (see e.g. \cite{Pol});
\be
\log Z^s(M) = Z^f(M). \label{fsrelation}
\ee
 Then we especially take a free field theory on the $D$ dimensional flat space $M=R^{D}$, whose coordinate denoted by $(x_0,x_1,\ddd,x_{D-1})$. Here we consider an Euclidean continuation such that $x_0$ is the Euclidean time. We define the subsystem $A$ to be $x_1>0$.  Then combining $(x_0,x_1)$ into a complex plane $C$, we can introduce the $n$-th R\'enyi entanglement entropy (REE) , denoted by $S^{(n)}_A$, via an analytic continuation $n=1/N$ from the orbifolds $C/\mathbb{Z}_N$
 as follows (refer to \cite{NiTa})
\be
S^{(n)}_A=\f{1}{1-n}\left[Z^f(C/\mathbb{Z}_N\times R^{D -2})-\f{1}{N}Z^f(C\times R^{D -2})\right]\biggr|_{N=\f{1}{n}}.
\label{fst}
\ee The $\mathbb{Z}_N$ orbifold action $g$ is given by
\be
g: (X,\bar{X})\to (e^{\f{2\pi i}{N}}X,e^{\f{-2\pi i}{N}}\bar{X}). \label{projgd}
\ee

\subsection{Entanglement Entropy for Free Scalar}

First we consider the EE for a free massive scalar field. The partition function of a spinless particle on the flat space is given by
\ba
Z^f(R^{D}) = \int _{\ep^2}^{\infty} \frac{d s}{2s } \Tr \ e^{ - s (\hat{k} ^2 + m^2 )}  &=& \frac{V_{D}}{(2 \pi)^{D}} \int _{\ep^2 }^{\infty} \frac{ds}{2s}
\int d^{D} k\ e^{-s (k^2 + m^2 )} \notag \\
&=& V_{D} \int ^{\infty}_{\ep^2 } \frac{ds}{2 s} (4 \pi  s)^{-\frac{D}{2}} e^{ - s m^2},
\ea
where the parameter $s $ is the Schwinger parameter which can be see as a moduli in the first quantization approach. Here, the trace is taken in the Hilbert space of one particle quantum mechanics and we choose the momentum eigenfunction $\ket{\vec{k}}$ as the base.
To obtain the partition function on $\mathbb{Z}_N$ orbifold , we need to take the trace over the subspace symmetric under the orbifold action. This can be done by inserting the projection
operator;
\ba
Z^f(C / \mathbb{Z}_N \times R^{D - 2}) &=& \int _{\ep^2}^{\infty} \frac{d s}{2s} \Tr \ \frac{1}{N} \sum_{j = 0}^{N -1} g^j \ e^{ - s (\hat{k }^2 + m^2 )} \notag \\
&=& \int _{\ep^2 }^{\infty} \frac{ds}{2s}
\int d^{D} k \ \frac{1}{N} \sum_{j = 0}^{N-1} \bra{\vec{k}} g^j\ket{\vec{k}} e^{-s (k^2 + m^2 )} .
\ea
Here $g$ is the generator of $\mathbb{Z}_N$. We can evaluate $\bra{\vec{k}} g^j \ket{\vec{k}}$ for $j \neq 0$ as follows:
\be
\bra{\vec{k}} g^j \ket{\vec{k}}= \delta^{D} (\vec{k} - g^j\cdot \vec{k}) = \frac{V_{D-2}}{(2 \pi)^{D-2}} \frac{1}{4 \sin^2 \frac{\pi j}{N}} \delta(k_0) \delta(k_1).
\ee
If we use the formula
\be
\sum_{j= 1} ^{N-1}  \frac{1}{\sin ^2 \frac{\pi j}{N}} = \frac{N^2 -1 }{3}, \label{bosonsum}
\ee
we get the final expression for the REE;
\be
S^{(n)} _A= \frac{(n+1) \pi V_{D-2}}{6 n} \int _{\ep ^2 }^{\infty} \frac{ds}{(4 \pi s )^{\frac{D}{2}}}\  e ^{- m^2 s } .
\ee
Especially the coefficient $V_{D-2}$ shows the area law of REE.

\subsection{Analysis of Bosonic Higher Spin Fields}

Next, we consider the generalization to other free fields with higher spins. First, we consider the bosonic fields. If the particle is fermion, a further modification is needed compared to the case of bosons as we will discuss in the next subsection.
 To get the representation for bosonic fields with any spins, we need to introduce the internal space of spin which corresponds to the components of the field.
 From this the base of Hilbert space is changed from $\ket{\vec{k}}$ to $\ket{\vec{k}, a} = \ket{\vec{k}} \otimes \ket{a}$ where $\{\ket{a}\}$ is the basis of the internal space.
In this case the internal space doesn't affect the energy of particle but the orbifold group $\mathbb{Z}_N$ acts on this space and thus the summation (\ref{bosonsum}) should be changed.
The final expression is given by
\ba
Z^f(C / \mathbb{Z}_N \times R^{D-2}) &=&  \int _{\ep^2}^{\infty} \frac{d s}{2 s} \Tr \ \frac{1}{N} \sum_{j = 0}^{N -1} g^j \ e^{ -  s(\hat{k }^2 + m^2 )}\notag \\
  &=&  \int _{\ep^2 }^{\infty} \frac{ds}{2s}
\int d^{D} k \ \frac{1}{N} \sum_{j = 0}^{N-1} \sum_{a =1}^{N_a}  \bra{\vec{k} ,a} g^j\ket{\vec{k},a} e^{-s (k^2 + m^2 )} .
\ea
where $N_a $ is the number of components and $F$ is the fermion number. $\braket{\vec{k},a |g^j|\vec{k},a}$ can be evaluated for $j \neq 0$  as follows:
\be
\braket{\vec{k},a|g^j|\vec{k},a} = \frac{V_{D-2}}{ (2 \pi)^{D-2} } \frac{e^{\frac{2 \pi i j s_a}{N}}}{4 \sin ^2 \frac{\pi j}{N}}\delta(k_0)\delta(k_1) .\label{bosonspin}
\ee
Here $s_a $ is the spin of $SO(2) \subset SO(D)$ which is the rotation of $(x_0,x_1)$ plane.

In this way we find that the basic sum formula which we can employ to calculate the entanglement entropy for free bosons is
\be
I(r,N)=\sum_{\beta=1}^{N-1}
\f{\cos\left(\f{2\pi \beta r}{N}\right)}{\sin^2\left(\f{\pi\beta}{N}\right)}
=\f{1}{3}\left[N^2-1+6N^2\left(\left\{\f{r}{N}\right\}^2-\left\{\f{r}{N}\right\}\right)\right], \label{bosoniccase}
\ee
where $\{x\}$ denotes the fractional part of $x$ i.e. $\{x\}=x-[x]$. Here we defined an integer $r$ by
\be
r=s_a.  \label{rsaa}
\ee

\subsection{Analysis of Fermionic Higher Spin Fields}

If the particle is a fermion, we need an overall minus sign because (\ref{fsrelation}) needs minus sign in front of $Z^f$ which follows from fermionic functional determinant.
Also $N$ should be an odd number in the case of fermions \cite{TaUe}. This is because the spin $s_a$ is a half integer and thus $g^N \neq 1$, while we have $g^{2N}=1$. Therefore we cannot regard it as a $\mathbb{Z}_N$ orbifold for even $N$. On the other hand, for $N$ odd, we obtain a $\mathbb{Z}_N$ orbifold by choosing $g^2$ as the generator of $\mathbb{Z}_N$ action, even in the presence of fermions. In this way, we find the final expression, which can be applicable to both a free boson and fermion, as follows:
\ba
Z^f(C / \mathbb{Z}_N \times R^{D-2}) &=& (-1 )^F \int _{\ep^2}^{\infty} \frac{d s}{2 s} \Tr \ \frac{1}{N} \sum_{j = 0}^{N -1} g^{2j} \ e^{ -  s(\hat{k }^2 + m^2 )}\notag \\
  &=& (-1)^F \int _{\ep^2 }^{\infty} \frac{ds}{2s}
\int d^{D} k \ \frac{1}{N} \sum_{j = 0}^{N-1} \sum_{a =1}^{N_a}  \bra{\vec{k} ,a} g^{2j}\ket{\vec{k},a} e^{-s (k^2 + m^2 )} . \label{fermionodd}
\ea
where $N_a $ is the number of components and $F$ is the fermion number. $\braket{\vec{k},a | g^{2 j} | \vec{k} ,a} $ can be evaluated for $j \neq 0$ by changing $j $ in (\ref{bosonspin}) to $2j$ ;
\be
\braket{\vec{k},a | g^{2 j} | \vec{k} ,a} = \frac{V_{D-2}}{(2 \pi)^{D-2}} \frac{e^{ \frac{2\pi i r}{N} }}{ 4\sin ^2 \frac{2 \pi j}{N}} \delta(k_0)\delta(k_1).
\ee

%Thus we find its derivative at $N=1$:

%\be
%\f{\de I(r,N)}{\de N}\Bigr | _{N=1}=\f{2}{3}-2|r|. \label{bosonI}
%\ee

When we calculate entanglement entropy for fermion, we need the following formula:
\be
J(r,N)=\sum_{\beta=1}^{N-1}
\f{\cos\left(\f{2\pi \beta r}{N}\right)}{\sin^2\left(\f{2\pi\beta}{N}\right)},
\ee
where we set $r=2s_a$ as in (\ref{rsaa}).

We can show that this summation is given as follows. When $r$ is an odd integer (i.e. fermions)
\be
J(r,N)|_{r\in odd}=\f{1}{3}(N^2-1)+2N^2\left[\left\{\f{r+N}{2N}\right\}^2-\left\{\f{r+N}{2N}\right\}\right].
\label{xo}
\ee
On the other hand, when $r$ is an even integer (i.e. bosons), we find
\be
J(r,N)|_{r\in even}=\f{1}{3}(N^2-1)+2N^2\left[\left\{\f{r}{2N}\right\}^2-\left\{\f{r}{2N}\right\}\right].\label{xe}
\ee
One can confirm that in the bosonic case the formula is just the same with (\ref{bosoniccase}).  It is easy to confirm $J(r,1)=0$ for any $r$.

For a specific values of $(r,N)$,  we find
\ba
&& J(r,N)|_{r\in odd}=-\f{N^2}{6}+\f{r^2}{2}-\f{1}{3}\ \ \ (-N\leq r\leq N), \label{forla}\\
&& J(r,N)|_{r\in even}=\f{N^2}{3}-rN+\f{r^2}{2}-\f{1}{3}\ \ \ (0\leq r\leq 2N).  \label{forlb}
\ea
Note that $J(r,N)$ is analytic for all $N\geq 1$ only if $r=0,1,2$, where (\ref{forla}) and (\ref{forlb}) can always be applied. This leads to a subtle issue for fields with spins higher than $3/2$.

\subsection{Conical Entropy for Higher Spin Fields}

Because the sums (\ref{xo}) and (\ref{xe}) are not analytic functions of $N$ in general, we need to define a rule how to take the derivative with respect to $N$. Here we choose to take the derivative w.r.t $N$ by employing the formula (\ref{forla}) and (\ref{forlb}), which are true when $N$ is large, as if they were
correct even at $N=1$. An advantage is that we can deal with analytic functions though they do not satisfy $J(1,r)=0$.
%\ba
%&&J(r,N)|_{r\in odd} = \f{1}{3}(N^2-1)+2N^2\left[\left(\frac{r+N}{2N}\right )^2-\left(\f{r+N}{2N}\right)\right] \\
%&&J(r,N)|_{r\in even} = \f{1}{3}(N^2-1)+2N^2\left[\left(\frac{r}{2N}\right )^2-\left(\f{r}{2N}\right)\right].
%\ea
This leads to
\ba
&& \f{\de J(r,N)}{\de N}\Biggr|_{N=1}=-\f{1}{3}, \ \ (r\in odd)  \label{oddcase} \\
&& \f{\de J(r,N)}{\de N}\Biggr|_{N=1}=\f{2}{3}-|r| \ \ (r\in even).  \label{evencase}
\ea
As we will see below, this prescription correctly reproduces the independent calculations of conical entropy in \cite{Kab,FM} for spin $0,1/2,1,3/2$ and $2$.

Then, applying this formula to calculate the conical entropy of free field theories in
$D$ dimension, we obtain the following form
\be
S_A=c_{ent}\cdot V_{D-2}\int^\infty_{\ep^2}\f{ds}{2s(4\pi s)^{\f{D-2}{2}}}e^{-m^2 s}, \label{sahsd}
\ee
where the coefficient $c_{ent}$ is computed as follows
\ba
c_{ent}&=&\f{1}{4}(-1)^F \sum_{a}\f{\de}{\de N}\left[\f{J(2s_a,N)}{N}-\f{J(2s_a,1)}{N}\right]\Biggr|_{N=1} \no
&=&\f{1}{4}(-1)^F \sum_{a}\f{\de J(2s_a,N)}{\de N}\Biggr|_{N=1},
\ea
where $a$ runs all spin components of the field.

For fermions, $\f{\de J}{\de N}|_{N = 1}$ does not depend on the spin $s=r/2$ and we always get
\be
c^{Fermion}_{ent}=\f{1}{12}\cdot[\# \mbox{Majorana spin components}].
\ee
This agrees with the know results for spin $1/2$ \cite{Kab} and $3/2$ \cite{FM}.

On the other hand, for bosons, $\f{\de J}{\de N}|_{N = 1}$ depends on the spin and we get
\be
c^{Boson}_{ent}=\f{1}{6}N_{dof}-\f{1}{2}\sum_{a=1}^{N_{dof}}|s_a|,
\ee
where  $N_{dof}$ denotes the number of components i.e. the number of real bosons and the number of Majorana fermion component and $|s_a|$ denotes the $SO(2)$ spin of $a$ component.
For example, a real scalar has $N_{dof}=1$ and $s_1=0$, which leads to
\be
c_{ent}(\mbox{Scalar})=\f{1}{6}\cdot  [\# \mbox{real scalars}].
\ee

For a $D $ dimensional $U(1)$ gauge field (without ghosts), we have $N_{dof}=D$, $s_1=-s_2=1$ and $s_3=s_4=\ddd=s_{D}=0$, which leads to
\be
c_{ent}(\mbox{Gauge})=\f{1}{6}\cdot D-1.
\ee
This agrees with the result in \cite{Kab} (by removing the ghost contributions).

For the spin 2, which is a graviton field $g_{\mu\nu}$, we can count the contributions in the same way. We find the values of $SO(2)$ spin $s$ and the number of components as follows:
\ba
&& s=2:\  1,\no
&& s=-2:\  1,\no
&& s=1:\   D-2,\no
&& s=-1:\  D-2,\no
&& s=0:\  \f{D^2-3D+4}{2}.
\ea
Therefore the total contribution to the EE is found as
\be
c_{ent}(\mbox{Spin}2)= 2\cdot \left(\f{1}{6}-1\right)+2(D-2)\cdot \left(\f{1}{6}-\f{1}{2}\right)
+\f{D^2-3D+4}{2}\cdot \f{1}{6}=\f{1}{6}\cdot\f{D(D+1)}{2}-D.
\ee
This agrees with the BH entropy calculation in \cite{FM}.

To see the relation to \cite{FM} more directly (note $\beta=2\pi/N$ in \cite{FM}), let us remind that the coefficient $A^{(j)}_1$ for a spin $j$ field
is defined by the coefficient in the heat kernel expansion:
\be
\log Z^{(j)}=(-1)^F\int^\infty_{\ep^2}e^{-m^2 s}\cdot (4\pi s)^{-D/2}\left(A^{(j)}_0+sA^{(j)}_1+\ddd\right)\int_{R^{D-2}},
\ee
where $\int _{R^{D-2}}$ denotes the volume of $R^{D-2}$.
Then we find that $A^{(j)}_1$ behaves when $N\simeq 1$ as follows in the Taylor expansion of
the powers of $N-1$:
\be
A^{(j)}_1=Q^{(j)}+4\pi c_1^{(j)}(N-1)+O\left((N-1)^2\right), \label{surfaceterm}
\ee
where the constant $Q^{(j)}$ denotes a singular contribution, which is non-zero only for
$j\geq 3/2$. This leads to the conical entropy (we set $c_{ent}=(-1)^{2j}c^{(j)}_{ent}$ in (\ref{sahsd})):
\be
S_A=(-1)^F\cdot c^{(j)}_{ent}\cdot V_{D-2}\int^\infty_{\ep^2}\f{ds}{2s(4\pi s)^{\f{D-2}{2}}}e^{-m^2 s},
\ee
where $c^{(j)}_{ent}$ is given by
\be
c^{(j)}_{ent}=4\pi c^{(j)}_1+Q^{(j)}.
\ee
The term $Q^{(j)}$ looks like a somewhat topological contribution analogous to the boundary entropy \cite{AfLu,CalCa} and the topological entanglement entropy \cite{KiPre,LeWe}.

It is also easy to compare this definition to our summation $J(r,N)$ in the orbifold method:
\be
A^{(j)}_1=\pi\cdot\f{J^{(j)}_{tot}(N)}{N}.
\ee
Here  $J_{tot}(N)$ is the summation of $J(r,N)$ for all components:
\be
J^{(j)}_{tot}(N)=\sum_{a}J(2s_a,N),
\ee
where $s_a$ is $SO(2)$ spin for each component of the spin $j$ field.

For the spin 0 (scalar) and spin 1/2 (Majorana fermion) we find
\ba
&& J^{(0)}_{tot}(N)=J(0,N)=\f{N^2-1}{3},\no
&& J^{(1/2)}_{tot}(N)=J(1,N)\cdot N_f=-\f{N^2-1}{6}\cdot N_f,
\ea
where $N_f$ is the number of components of Majorana spinor.
This agrees with
\be
A^{(0)}_1=\f{\pi}{3N}(N^2-1),\ \ \ A^{(1/2)}_1=-\f{N_f}{2}\cdot A^{(0)}_1.
\ee

For the spin 1, 3/2 and 2 we find
\ba
&& \pi J^{(1)}_{tot}(N)=2\cdot \pi  J(2,N)+(D-2) \pi J(0,N)=N\left(D\cdot A^{(0)}_1+4\pi(1/N-1)\right),\no
&& \pi J^{(3/2)}_{tot}(N)=N_f\cdot \pi J(3,N)+N_f\cdot (D-1)\pi \cdot J(1,N)=N\left(-\f{N_f }{2}A^{(0)}_1 D+\frac{4\pi}{N} N_f\right),\no
&&\pi J^{(2)}_{tot}(N)=2 \pi J(4,N)+2(D-2)\pi J(2,N)+\f{D^2-3D+4}{4} \pi J(0,N) \no
&& =N\left(\f{D(D+1)}{2}\cdot A^{(0)}_1+4\pi(D+2)(1/N-1)+8\pi\right).
\ea
All of these agree with (2.10)-(2.13) in \cite{FM}. Notice that our derivation is much simpler as the calculation is essentially reduced to the evaluation of $J(r,N)$.

%Finally, if we use these our formula for closed string theory and evaluate $g(N)$ (\ref{gn}) we get
%\be
%g'(1)=\f{1}{32}.
%\ee
%This is the same value we used in section 5.3 .

\subsection{Thermal Entropy in Rindler Space for Higher Spin Fields}
In the previous subsection, we took the derivative of $J(r,N)$ assuming $N$ is sufficiently
large compared to $r$ and interpolated the result to $N=1$. If we take the derivative w.r.t $N$ by assuming $N$ is extremely large and by ignoring $r$ dependent term, we get
\ba
&& \f{\de J(r,N)}{\de N}\Biggr|_{N=1}=-\f{1}{3}, \ \ (r\in odd) \no
&& \f{\de J(r,N)}{\de N}\Biggr|_{N=1}=\f{2}{3} \ \ (r\in even).
\ea
In this calculation, there are no spin dependent terms and all fields contribute to entanglement entropy universally:
\ba
&&c_{ent}(\mbox{fermion}) = \frac{1}{12} \cdot [\# \mbox{Majorana spin components}  ]
\label{eewq}  \\
&&c_{ent}(\mbox{boson}) = \frac{1}{6} \cdot [\# \mbox{components}] \label{eeqq}
\ea
This calculation corresponds to ignoring the surface terms and picking up the thermodynamical entropy in the Rindler space.

\section{Conical Entropy in Open String}

In this section we will start our analysis of conical entropy in string theory by studying open string theories briefly.

\subsection{Conical Entropy in Open Bosonic String}

The Bosonic openstring vacuum amplitude in the orbifold theory $C/\mathbb{Z}_N \times R^{24}$ is given by (we subtract the $N=1$ result)
\ba
&& Z_{open}[C/\mathbb{Z}_N\times R^{24}]-\f{1}{N}Z_{open}[C\times R^{24}] \no
&&  =V_{24}\int \f{dt}{2t}(4\pi^2\al t)^{-12}
\sum_{\beta=1}^{N-1}\f{1}{N\sin\left(\f{\pi\beta}{N}\right)
\theta_1\left(\f{\beta}{N}|it\right)\eta(it)^{23}},
\ea
where the theta function is given by
\be
\theta_1(x|it)=2e^{-\f{\pi}{4}t}\sin(\pi x)\prod_{n=1}^\infty (1-e^{-2\pi tn})
(1-e^{-2\pi tn+2\pi ix})(1-e^{-2\pi tn-2\pi ix}).
\ee

Therefore we need to perform the following summation:
\be
I_{open}(N)=\sum_{\beta=1}^{N-1}
\f{1}{\sin\left(\f{\pi\beta}{N}\right)\theta_1\left(\f{\beta}{N}|it\right)}.
\ee
Then the conical entropy can be found as
\be
S_A=V_{24}\int \f{dt}{2t}(4\pi^2\al t)^{-12}\f{1}{\eta(it)^{23}}\cdot \f{\de I_{open}(N)}{\de N}\Biggr|_{N=1}, \label{opence}
\ee

By expanding $I_{open}(N)$ w.r.t. the powers of $q=e^{-2\pi t}$ we obtain
\be
I_{open}(N)=\f{1}{2}\cdot\f{e^{\pi t/4}}{\prod_{n=1}^\infty(1-e^{-2\pi nt})}\cdot
\sum_{\beta=1}^{N-1}\f{1}{{\sin^2\left(\f{\pi\beta}{N}\right)}}\prod_{n=1}^\infty \left[\sum_{p_n=0}^\infty \sum_{q_n=0}^\infty
e^{2\pi i\f{\beta}{N}(p_n-q_n)}\cdot e^{-2\pi nt(p_n+q_n)}\right].
\ee
Since $I_{open}$ is real valued we find
\be
I_{open}(N)=\f{1}{2}\cdot\f{e^{\pi t/4}}{\prod_{n=1}^\infty(1-e^{-2\pi nt})}\cdot
\sum_{\beta=1}^{N-1}\f{1}{{\sin^2\left(\f{\pi\beta}{N}\right)}} \left[\sum_{\vec{p}=0}^\infty \sum_{\vec{q}=0}^\infty
\cos\left(2\pi\f{\beta}{N}\sum_{n=1}^{\infty}(p_n-q_n)\right)\cdot e^{-2\pi t\sum_{n=1}^\infty n(p_n+q_n)}\right].
\ee

Now we take the derivative w.r.t $N$ and take the limit $N=1$ using the formula (\ref{evencase}) we obtain
\ba
&& \f{\de I_{open}(N)}{\de N}\Bigr | _{N=1} \no
&& =\f{1}{2}\cdot\f{e^{\pi t/4}}{\prod_{n=1}^\infty(1-e^{-2\pi nt})}\cdot
\left[\sum_{\vec{p}=0}^\infty \sum_{\vec{q}=0}^\infty \left(\f{2}{3}-2\left|\sum_{n=1}^{\infty}(p_n-q_n)\right|\right)
\cdot e^{-2\pi t\sum_{n=1}^\infty n(p_n+q_n)}\right].
\ea
Unfortunately it seems difficult to make the above expression simpler. However, it is clear that the conical entropy (\ref{opence}) has divergences due to the closed string tachyon.

\subsection{Conical Entropy in Open Superstring}

The superstring amplitude in the orbifold theory $C/\mathbb{Z}_N \times R^{8}$ is given by (we subtract the $N=1$ result)
\ba
&& Z_{open}[C/\mathbb{Z}_N\times R^{8}]-\f{1}{N}Z_{open}[C\times R^{8}] \no
&&  =V_{8}\int \f{dt}{2t}(4\pi^2\al t)^{-4}
\sum_{\beta=1}^{N-1}\f{\theta_1\left(\f{\beta}{N}|it\right)^4}
{N\sin\left(\f{2\pi\beta}{N}\right)\theta_1\left(\f{2\beta}{N}|it\right)\eta(it)^{9}}
\ea

Therefore we need to take the summation:
\be
J_{open}^{susy} = \sum_{\beta = 1} ^{N-1} \frac{\theta_1(\frac{\beta}{N} | it)^4 }{\sin (\frac{2 \pi \beta}{N}) \theta _1 (\frac{ 2 \beta}{N}|it)}
\ee
To evaluate conical entropy, we use Jacobi identity to separate the contribution
from bosons and that of fermions:
\be
\prod_{a=1} ^{4} \theta _3(\nu_a | \tau) -\prod_{a=1} ^{4} \theta _2(\nu_a | \tau)
- \prod_{a=1} ^{4} \theta _4(\nu_a | \tau) + \prod_{a=1} ^{4} \theta _1(\nu_a | \tau)
= 2 \prod_{a = 1}^4\theta _1(\nu '_a|\tau).
\ee
Here we have defined
\ba
2 \nu'_1 = \nu_1 + \nu _2 + \nu_3 + \nu_4, \ \ \ 2 \nu'_2 = \nu_1 + \nu _2 - \nu_3 - \nu_4 \\
2 \nu'_3 = \nu_1 - \nu _2 + \nu_3 - \nu_4, \ \ \ 2 \nu'_4 = \nu_1 - \nu _2 - \nu_3 + \nu_4
\ea
and $\theta _2(\nu|\tau), \theta _3(\nu|\tau),\theta _4(\nu|\tau)$ is given by
\ba
&&\theta _2(\nu, \tau) = \sum_{n = -\infty}^{\infty} q^{(n - \frac{1}{2})^2}  z ^{n - \frac{1}{2}} \\
&&\theta _3(\nu, \tau) = \sum_{n = -\infty}^{\infty} q^{n^2}  z ^{n } \\
&&\theta _4(\nu, \tau) = \sum_{n = -\infty}^{\infty} (-1)^n q^{n^2}  z ^{n }
\ea
where $q = e^{2 \pi i \tau}$ and $z = e^{2 \pi i \nu}$. Using these formula and expanding the powers of $q = e^{-2 \pi t}$
$J_{open}^{suzy}$ becomes as follows
\ba
&&J_{open}^{susy} = \frac{1}{4} \frac{e^{\pi t /4}}{ \prod_{n=1}^{\infty} (1 - e^{- 2 \pi n t})} \cdot \sum_{\beta = 1} ^{N -1} \frac{1}{\sin ^2 (\frac{\pi \beta}{N})}
\notag  \\
&& \ \ \  \times \Bigg[ \theta_3 (0|it)^3 \sum_{m \in \mathbb{Z}} \sum_{\vec{p} = 0}^{\infty} \sum_{\vec{q}=0}^{\infty} \cos \Bigg( 2 \pi \frac{\beta}{N} \sum_{n=1}^{\infty} (p_n - q_n + m) \Bigg) \cdot e^{- 2 \pi t (\sum_{n = 1 }^{\infty} n(p_n + q_n)  + m^2 /2 ) } \notag \\
&& \ \ \ -\theta_4 (0|it)^3 \sum_{m \in \mathbb{Z}} \sum_{\vec{p} = 0}^{\infty} \sum_{\vec{q}=0}^{\infty} (-1) ^m\cos \Bigg( 2 \pi \frac{\beta}{N} \sum_{n=1}^{\infty} (p_n - q_n + m) \Bigg) \cdot e^{- 2 \pi t (\sum_{n = 1 }^{\infty} n(p_n + q_n)  + m^2 /2 ) }
\notag \\
&& \ \ \ -\theta_2 (0|it)^3 \sum_{m \in \mathbb{Z}} \sum_{\vec{p} = 0}^{\infty} \sum_{\vec{q}=0}^{\infty} \cos \Bigg( 2 \pi \frac{\beta}{N} \sum_{n=1}^{\infty} (p_n - q_n + m - \frac{1}{2}) \Bigg) \cdot e^{- 2 \pi t (\sum_{n = 1 }^{\infty} n(p_n + q_n)  + (m - \frac{1}{2})^2 /2 ) } \Bigg] . \notag  \\
\ea
Now we take the derivative w.r.t $N$ and take the limit $N =1$ using the formula (\ref{oddcase}) (\ref{evencase})  we obtain
\ba
&&\frac{\partial J_{open}^{susy}}{\partial N} \Big| _{N  =1} = \frac{1}{4} \frac{e^{\pi t /4}}{ \prod_{n=1}^{\infty} (1 - e^{- 2 \pi n t})} \cdot \sum_{\beta = 1} ^{N -1} \frac{1}{\sin ^2 (\frac{\pi \beta}{N})}
\notag  \\
&& \ \ \  \times \Bigg[ \theta_3 (0|it)^3 \sum_{m \in \mathbb{Z}} \sum_{\vec{p} = 0}^{\infty} \sum_{\vec{q}=0}^{\infty}  \Bigg( \frac{2}{3} - \Bigg | \sum_{n=1}^{\infty} (p_n - q_n + m)  \Bigg |\Bigg) \cdot e^{- 2 \pi t (\sum_{n = 1 }^{\infty} n(p_n + q_n)  + m^2 /2 ) } \notag \\
&& \ \ \ -\theta_4 (0|it)^3 \sum_{m \in \mathbb{Z}} \sum_{\vec{p} = 0}^{\infty} \sum_{\vec{q}=0}^{\infty} (-1) ^m \Bigg( \frac{2}{3} - \Bigg | \sum_{n=1}^{\infty} (p_n - q_n + m)  \Bigg |\Bigg) \cdot e^{- 2 \pi t (\sum_{n = 1 }^{\infty} n(p_n + q_n)  + m^2 /2 ) }
\notag \\
&& \ \ \  + \frac{1}{3}\theta_2 (0|it)^3 \sum_{m \in \mathbb{Z}} \sum_{\vec{p} = 0}^{\infty} \sum_{\vec{q}=0}^{\infty}  e^{- 2 \pi t (\sum_{n = 1 }^{\infty} n(p_n + q_n)  + (m - \frac{1}{2})^2 /2 ) } \Bigg] \notag . \\
\ea

This results is not vanishing unless some miraculous cancellations occur, though the superstring vacuum amplitude is zero.
This is because the orbifold does not preserve supersymmetry and the partition function
does not become $0$ if $N \neq 1$. Moreover, we find that conical entropy in open superstring has the UV divergence due to the $t$ integral. This is simply understood as follows. The above evaluation can be regarded as a simple summation of conical entropy for all higher spin fields and thus it is a sum of the area law divergences. This fact that the open string conical entropy is divergent shows that we need to take into account the
backreaction of open string sectors to closed string sectors. We expect that we can get a finite conical entropy if we can treat this backreaction properly, though this is beyond the scope of this paper. This issue motivates us to study the conical entropy in closed string.

\section{(Twisted) Conical Entropy in Closed Superstring}

In this section we study the main problem of this paper: conical entropy in type II closed superstring. We will focus on the contributions from the torus amplitude, which lead to the leading quantum corrections, while the contribution from the sphere amplitude are expected to lead to the Bekenstein-Hawking formula \cite{SU}.
We will start from a definition of conical entropy. Then we will introduce a twisted conical entropy, which is easier to evaluate and we will analyze this quantity in detail.

\subsection{Definition of Conical Entropy in Closed Superstring}

We express the partition function of closed string theory on the spacetime $M$ at the first quantized level as $Z_{closed}(M)$. We especially focus on type II string theory on the flat space $M=R^{10}$, whose coordinate denoted by $(x_0,x_1,\ddd,x_9)$. We define the subsystem $A$ to be $x_1>0$.  Then combining $(x_0,x_1)$ into a complex plane $C$, we can introduce the $n$-th Renyi entanglement entropy (EE) of string theory, denoted by $S^{(n)}_A$, as follows
\be
S^{(n)}_A=\f{1}{1-n}\left[Z_{closed}(C/\mathbb{Z}_N\times R^8)-\f{1}{N}Z_{closed}(C\times R^8)\right]\biggr|_{N=\f{1}{n}},
\label{fsst}
\ee
where $C/\mathbb{Z}_N$ is the standard $\mathbb{Z}_N$ orbifold in type II string. The $\mathbb{Z}_N$ orbifold action $g$ is given by
\be
g: (X,\bar{X})\to (e^{\f{2\pi ik}{N}}X,e^{\f{-2\pi ik}{N}}\bar{X}),
\ee
where $k$ is a positive integer fixed below.
We are especially interested in $n=1$ (i.e. $N=1$) limit, being equivalent to the von-Neumann entropy. In order to have $\mathbb{Z}_N$ orbifold in type II string, $N$ should be an odd integer and $k$ should be an even integer \cite{APS,TaUe}. Thus we will set $k=2$ without losing generality.

The partition function of type II string on $C/\mathbb{Z}_N\times R^8$ is given by
(refer to \cite{LS,SU,Emp,Dah1,Dah2} for earlier works)
\ba
&& Z_{closed}(C/\mathbb{Z}_N\times R^8) \no
&& =V_8\int_F\f{d\tau^2}{4\tau_2}\cdot (4\pi^2\al\tau_2)^{-4}
\cdot\sum_{l,m=0}^{N-1}\f{|\theta_1(\nu_{lm}/2|\tau)|^8}{N|\eta(\tau)|^{18}
|\theta_1(\nu_{lm}|\tau)|^2}, \label{orbamp}
\ea
where $\nu_{lm}=\f{k(l-m\tau)}{N}$.

\subsection{Twisted conical entropy from Melvin background}

If we want to evaluate the conical entropy directly from the $C/\mathbb{Z}_N$ orbifold amplitude (\ref{orbamp}), we need to perform the two summations with respect to
 $l$ and $m$. Since this looks rather hard, we would like to focus on a modified quantity which can be computed in an easier way.

For this purpose, we would like to consider superstring on so called Melvin background
defined in \cite{Tsy,Bosonic,RuTs}. Though there are several ways to introduce Melvin backgrounds, the most simple one which fits nicely with our purpose is the one defined as
a $\mathbb{Z}_N$ orbifold (or called a twisted circle):
\be
\mbox{Melvin background}: (C\times S^1)/\mathbb{Z}_N\times R^7,  \label{orbimel}
\ee
where the radius of the circle $S^1$ before the $\mathbb{Z}_N$ orbifold is defined to be
$NR$. In the above, the $\mathbb{Z}_N$ orbifold action $g$ is defined by
\be
g: (X,\bar{X},y)\to (e^{\f{2\pi ik}{N}}X,e^{\f{-2\pi ik}{N}}\bar{X},y+2\pi R),
\ee
where $R$ is the radius of the Melvin background. Again in type II string, $k$ is an even integer, while $N$ is an odd integer. We can set $k=2$ without losing generality.

We can explicitly show that the Melvin background (\ref{orbimel}) is reduced to the original orbifold $C/\mathbb{Z}_N\times \ti{S}^1\times  R^7$ \cite{TaUe,brane} if we take the small radius limit $R\to 0$ using the T-duality. The radius of the T-dualized circle $\ti{S}^1$ is given by $R_{orb}=\f{\al}{NR}$.

Now we define the twisted conical entropy as follows:
\be
\tilde{S}^{(n)}_A=\f{1}{1-n}\left[Z_{closed}\left[(C\times S^1)/\mathbb{Z}_N\times R^7\right]-\f{1}{N}Z_{closed}\left[C\times S^1\times R^7\right]\right]\Biggl|_{n=1/N}.
\ee

The von-Neumann entropy limit $N=1$ can be computed as
\ba
\ti{S}_A& \equiv &\ti{S}^{(1)}_A = Z_{closed}\left[C\times S^1\times R^7\right]+\f{\de}{\de N}Z_{closed}\left[(C\times S^1)/\mathbb{Z}_N\times R^7\right]\Bigl |_{N=1},\label{partmek}
\ea
where the first term in the right hand side is vanishing due to the supersymmetries.
Note that we choose the radius of $S^1$ (before the orbifold projection) in $(C\times S^1)/\mathbb{Z}_N\times R^7$ to be $NR=\f{\al}{R_{orb}}$, while that in $C\times S^1 \times R^7$ to be $R=\f{\al}{R_{orb}}$. When we take the derivative w.r.t $N$ in (\ref{partmek}),
we keep $R_{orb}$ fixed, motivated by the original definition (\ref{fsst}).
Based on this definition, we will work out the twisted conical entropy from the partition function in Melvin model in the coming subsections.

\subsection{Partition function in Melvin Model and conical entropy}

The partition function of Melvin model \cite{RuTs} is given by
\be
Z_{closed}\left[(C\times S^1)/\mathbb{Z}_N\times R^7\right]=Z_0\cdot \int_F\f{d\tau^2}{\tau_2^5}\sum_{w',w=-\infty}^{\infty}
e^{-\f{\pi R^2}{\al \tau_2}|w-w'\tau|^2}\cdot \f{|\theta_1((w-w'\tau)/N|\tau)|^8}
{|\eta(\tau)|^{18}|\theta_1(2(w-w'\tau)/N|\tau)|^2}, \label{melmel}
\ee
where $Z_0=\f{V_7R}{4(2\pi)^7\ap'^4}$. The region $F$ represents the standard fundamental region of the torus moduli space. After some algebras we can easily see that for a large
radius $R$, there are no closed string tachyons, while for a small $R$ there exist tachyons in the twisted sectors \cite{RuTs,DGHM} so that in the small radius limit $R\to 0$ the theory is reduced to that for the orbifold $C/{\mathbb{Z}}_N\times R^8$ \cite{TaUe,brane}.

By using the folding procedure given in the appendix A, we can equivalently replace one of the two summations in the partition function as the summation over integration domains, which are obtained by the $SL(2,{\mathbb{Z}})$ modular transformations of the fundamental region $F$. This enables us to rewrite it as the integral over the strip $S$ defined by $-1/2<\tau_1<1/2$ and $\tau_2>0$, with a single sum:
\be
Z_{closed}\left[(C\times S^1)/\mathbb{Z}_N\times R^7\right]=Z_0\int_{S}\f{d\tau^2}{\tau_2^5}\sum_{w=-\infty}^\infty e^{-\f{\pi R^2}{\al \tau_2}w^2}\cdot
\f{|\theta_1(w/N|\tau)|^8}
{|\eta(\tau)|^{18}\cdot |\theta_1(2w/N|\tau)|^2},  \label{patmlsk}
\ee
where the integral region $S$ denotes the strip defined by $-1/2<\tau_1<1/2$ and $\tau_2>0$.

In order to obtain the twisted conical entropy, we can decompose the summation over $w$ as
\be
w=N\ap+\beta,
\ee
where $\ap$ runs all integers from $-\infty$ to $\infty$, while $\beta$ takes $0,1,2,\ddd,N-1$.

After we did the Poisson resummation:
\be
\sum_{\gamma\in {\mathbb{Z}}}\exp(-\pi a\gamma^2+2\pi ib\gamma)
=\f{1}{\s{a}}\sum_{\alpha\in {\mathbb{Z}}}\exp(-\f{\pi(\alpha-b)^2}{a}),
\ee
 we find
\be
Z_{closed}\left[(C\times S^1)/\mathbb{Z}_N\times R^7\right]=Z_0\int_{S}\f{d\tau^2}{\tau_2^5}\f{\s{\al\tau_2}}{NR}\sum_{\gamma\in {\mathbb{Z}}}\sum_{\beta=0}^{N-1}
e^{-\f{\pi \al\tau_2}{R^2N^2}\gamma^2}\cdot e^{2\pi i\f{\beta \gamma}{N}}
\f{|\theta_1(\beta/N|\tau)|^8}
{|\eta(\tau)|^{18}\cdot |\theta_1(2\beta/N|\tau)|^2}. \label{pos}
\ee
Note that we can omit the contribution from $\beta=0$ in (\ref{pos}) as it is vanishing using the properties of theta functions.

To simplify our analysis, we will take the two differen limits of integrand: the IR limit: $\tau_2\to\infty$ and UV limit: $\tau_2\to 0$. Then we will obtain some analytical results from these formula and we can obtain important behaviors of twisted conical entropy in the coming two subsections.

\subsection{IR limit $\tau_2\to\infty$}

 First let us study the expression (\ref{pos}) in the IR limit $\tau_2\to\infty$. In appendix B, we have shown the details how to
do the summation to the all orders of $\exp({-\tau_2})$. In this case, the summation is localized at $\gamma=0$. Thus we find in this limit:
\be
Z_{closed}\left[(C\times S^1)/\mathbb{Z}_N\times R^7\right]\simeq 64\cdot Z_0\int_{S}\f{d\tau^2}{\tau_2^5}\f{\s{\al\tau_2}}{NR}\cdot
\sum_{\beta=0}^{N-1}\f{\sin^8\left(\f{\pi \beta}{N}\right)}{\sin^2\left(\f{2\pi \beta}{N}\right)}.
\ee
Notice that there is no tachyonic divergence as the twisted sector contributions, which include localized closed string tachyons \cite{Dah1,Dah2,LS,APS}, are removed by the folding trick and are hidden in the $\tau_2\to 0$ limit as we will see in the next subsection.

Therefore we need to evaluate
\be
g(N)\equiv \sum_{\beta=0}^{N-1}\f{\sin^8\left(\f{\pi \beta}{N}\right)}{\sin^2\left(\f{2\pi \beta}{N}\right)}. \label{gn}
\ee

It is easy to see that for $N=3,5,7,\ddd$, we can show
\be
g(N)=\f{1}{4}N^2-\f{15}{32}N. \label{fgfhn}
\ee
Note that this does not satisfy the standard condition $g(1)=0$. This is the subtle issue discussed in the section 2.3. Following the prescription (\ref{oddcase}) and (\ref{evencase}) of conical entropy given in section 2.4, we can simply take the derivative of $g(N)$ in (\ref{fgfhn}) and set $N=1$, which leads to $g'(1)=1/32$. Indeed, if we decompose (\ref{gn}) into each $SO(2)$ spin components, they correspond to the type II supergravity multiplet.

In this way, we finally get the IR estimation of the twisted conical entropy $\ti{S}_A$ for the type II string theory on $R^9\times S^1$
(radius of $S^1$ is $R_{orb}$):
\be
\ti{S}_A(R_{orb})\simeq 2Z_0\cdot \f{R_{orb}}{\s{\al}}\int_S~ \f{d\tau^2}{\tau_2^{9/2}}
\sim \f{V_7}{\al^{7/2}},
\ee
which clearly converges in the integral region of the IR limit $\tau_2\to\infty$. That is to say there is no divergence induced by the tachyon in the (\ref{pos}) and the twisted conical entropy is well defined in the IR region.

\subsection{UV limit: $\tau_2\to 0$}

Next we would like to study expression (\ref{pos}) in the UV limit $\tau_2\to 0$.
Let us define the following function first\footnote{The precise study on the asymptotic behavior of negative index Jacobi forms $f(\tau)$ is given by residues theorem and Fourier expansion in \cite{Kathrin}.}:
\be
f(\tau)\equiv \f{\theta_1\left(\f{\beta}{N}|\tau\right)^4}
{\eta(\tau)^9\theta_1\left(\f{2\beta}{N}|\tau\right)}.
\ee
We can expand $f(\tau)$ as
\be
f(\tau)=\sum_{n=0}^\infty d_n e^{2\pi i\tau n}.
\ee
We want to make use
of the saddle point approximation to estimate the leading behavior of $\tau\to 0$.
By taking the limit $\tau\to 0$ and using the modular transformation, we find the behavior
\ba
f(\tau)&\simeq & (\tau_2)^3 e^{\f{2\pi\beta}{N\tau_2}} \ \ \ (0<\beta/N< 1/2),\no
&\simeq & (\tau_2)^3 e^{\f{2\pi}{\tau_2}\left(1-\f{\beta}{N}\right)} \ \ \ (1/2<\beta/N< 1).\no
\ea

This leads to the following estimation for large $n$ when $0<\beta/N<1/2$ (up to a numerical
constant):
\be
d_n\sim \left(\f{\beta}{N}\right)^{7/4} n^{-9/4}e^{4\pi\s{\f{\beta n}{N}}}, 
\ee
where we approximated as follows in the limit $\tau_2 \to 0$
\be
\sum_{n=0}^\infty e^{-2\pi\tau_2\left(\s{n}-\f{\s{\beta}}{\tau_2\s{N}}\right)^2}\simeq 2\int^\infty_{-\infty} dy \left(y+\f{\s{\beta}}{{\tau_2\s{N}}}\right)e^{-2\pi\tau_2 y^2}
=\s{\f{2\beta}{N}}(\tau_2)^{-3/2},
\ee
where we defined $y=\s{n}-\f{\s{\beta}}{\tau_2\s{N}}$. Thus we can evaluate the following important ingredient of our partition function
\be
\int^{1/2}_{-1/2}d\tau_1 |f(\tau)|^2=\sum_{n}(d_n)^2 e^{-4\pi\tau_2 n}\sim \s{\f{N}{\beta}}(\tau_2)^{15/2}\cdot e^{4\pi \f{\beta}{N\tau_2}}. \label{hage}
\ee
We can similarly obtain the estimation of $d_n$ when $1/2<\beta/N<1$ by replacing $\beta$ with
$N-\beta$.

 It is clear from (\ref{patmlsk}) that the divergences in the integration of $\tau_2$ near $\tau_2=0$ in the partition function are convergent for sufficiently large values of $R$. This means that the twisted (Renyi) conical entropy $\ti{S}^{(n)}_A$ is well-defined for any $N(\geq 1)$ when $R$ is large enough.

On the other hand, we are more interested in the small $R$ limit so that it is related to the original orbifold theory. Then we immediately find that the $\tau_2$ integral gets divergent near $\tau_2=0$ due to the Hagedorn behavior (\ref{hage}), which is related to the presence of closed string tachyons. However, as we will see below, we are still able to
evaluate the $N=1$ limit i.e. the twisted conical entropy $\ti{S}_A$, while for any $N>1$,
$\ti{S}^{(n)}_A$ suffers from the tachyonic divergences. However, note that the analytical continuation of the summation (\ref{funch}) does not have any divergence in the limit $\tau_2\to 0$ for the original Renyi entropy region $N=1/n<1$.

Now, in order to evaluate the partition function (\ref{pos}), we need to perform the summation over $\beta$:
\ba
h(N)&=&\sum_{\beta=1}^{\f{N-1}{2}}\s{\f{N}{\beta}}e^{2\pi i\f{\beta \gamma}{N}}e^{\f{4\pi\beta}{N\tau_2}}
+\sum_{\beta=\f{N+1}{2}}^{N-1}\s{\f{N}{N-\beta}}e^{2\pi i\f{\beta \gamma}{N}}e^{\f{4\pi(N-\beta)}{N\tau_2}}\no
&=&2~\mbox{Re}\left[\sum_{\beta=1}^{\f{N-1}{2}}\s{\f{N}{\beta}}e^{2\pi i\f{\beta \gamma}{N}}e^{\f{4\pi\beta}{N\tau_2}}\right], \label{funch}
\ea
where Re$[z]$ denotes the real part of $z$.
By replacing $\s{\f{N}{\beta}}$ with the integral $\int^\infty_{-\infty}dx e^{-\pi\beta x^2/N}$, we find
\be
\f{\de h(N)}{\de N}\Bigr|_{N=1}=\int^{\infty}_{-\infty}dx
\f{\f{4\pi}{\tau_2}-\pi x^2}{1-e^{-\f{4\pi}{\tau_2}+\pi x^2}},
\ee
where the $\gamma$ dependence cancels out.
This can be estimated in the limit $\tau_2\to 0$ as follows
\be
\f{\de h(N)}{\de N}\Bigr|_{N=1}=-\int^\infty_{-\f{4\pi}{\tau_2}}
\f{zdz}{\s{\pi}\s{\f{4\pi}{\tau_2}+z}(1-e^z)}\simeq \f{32\pi}{3} (\tau_2)^{-3/2},
\ee
where we defined $z=\pi x^2-\f{4\pi}{\tau_2}$.

The partition function (\ref{pos}) is now evaluated as follows (up to a $O(1)$ constant)
\be
Z_{closed}\left[(C\times S^1)/\mathbb{Z}_N\times R^7\right]\sim Z_0\cdot\f{R_{orb}}{\s{\al}}\int^{\tau_{max}}_{0}d\tau_2(\tau_2)^{3}\cdot h(N)\cdot\sum_{\gamma\in {\mathbb{Z}}}e^{-\f{\pi R^2_{orb}\tau_2\gamma^2}{\al}}, \label{wer}
\ee
where $\tau_{max}$ is the upper bound where we can apply the $\tau_2\to 0$ approximation.
Then, using the above results, we would like to estimate the twisted cone entropy by taking the derivative w.r.t $N$, following the definition (\ref{partmek}). This leads to (note
the identity $h(1)=0$)
\be
\ti{S}_A\sim  \f{V_7}{\al^{7/2}}\int^{\tau_{max}}_{0}d\tau_2(\tau_2)^{3/2}\sum_{\gamma\in {\mathbb{Z}}}e^{-\f{\pi R^2_{orb}\tau_2\gamma^2}{\al}}.
\ee
In this way, we finally obtain the following estimation for the UV contributions:
\be
[\ti{S}_A]_{UV}\simeq \f{V_7}{\al^{7/2}}\left(s_1+s_2\cdot \f{\al^{5/2}}{R_{orb}^5}\right), \label{xqx}
\ee
where $s_1$ and $s_2$ are $O(1)$ numerical constants and we assumed $R_{orb}\gg\s{\al}$. If we consider the opposite limit $R_{orb}\ll\s{\al}$,  we have (\ref{xqx}) with $s_2=0$.

\subsection{Summary}

In summary, we find that the moduli integral of the twisted conical entropy $\ti{S}_A$ does converge both in the IR and UV region. Since it is clear that these regions are only possible sources of divergence as usual in string theory, we can argue that the twisted conical entropy is finite in superstring. Note that if we do the similar analysis for bosonic string Melvin backgrounds \cite{Bosonic}, the twisted conical entropy will get divergent as is so in the torus partition function for bosonic string on the flat space.
Also the Renyi version $\ti{S}^{(n)}_A$ with $n=1/N<1$ turns out to be divergent even for superstring in general due to the closed string tachyons.

Our estimation leads to the following form of the (von-Neumann) twisted conical entropy in superstring on $C\times S^1\times R^7$:
\be
\ti{S}_A(R_{orb})=\f{V_7}{\al^{7/2}}\cdot \ti{S}\left(\f{R_{orb}}{\s{\al}}\right), \label{entrfinh}
\ee
where $V_7$ is the volume of the 7 dimensional flat space $R^7$.
Moreover, our analysis shows that the function $\ti{S}\left(\f{R_{orb}}{\s{\al}}\right)$ in  (\ref{entrfinh}) approaches
to a finite constant $\ti{S}_0$ in the limit $R_{orb}\to \infty$ as follows from (\ref{xqx}).
Therefore in the large radius limit $R_{orb}\to \infty$, the twisted conical entropy behaves as follows:
\be\label{law}
\ti{S}_A(R_{orb})\simeq \ti{s}\cdot \f{V_7}{\al^{7/2}}.
\ee
 If we do the same calculation in a quantum field theory compactified on a circle with the small radius $R=\f{\al}{R_{orb}}$, it is obvious that we will find it is
UV divergent as $\ti{S}_A\sim \f{V_7}{\vep^7}$, where $\vep$ is the UV cut off (or lattice spacing) because the Kaluza-Klein modes are suppressed in the small radius limit. In superstring, as we have seen, this UV divergence is removed owing to the string scale cutoff as expected. In this sense the above result is analogous to our original expectation (\ref{areastring}). 

Now we would like to get back to our original problem of computing the conical entropy $S_A$ in superstring. As the Melvin background $(C\times S^1)/\mathbb{Z}_N\times R^7$ is reduced to the $C/\mathbb{Z}_N\times R^8$ orbifold in the small radius limit $R=\f{\al}{NR_{orb}}\to 0$, we expect that the leading contribution of the twisted conical entropy $\ti{S}_A$ coincides with the conical entropy $S_A$ in this limit. It is natural to expect that the conical entropy $S_A$ in superstring is proportional to the 8 dimensional area of the boundary $\de A$ of the subsystem $A$ as in (\ref{areastring}). Therefore, the absence of a term proportional to the 8 dimensional area $V_7\cdot R_{orb}$ in $\ti{S}_A$ (\ref{law}) implies that the conical entropy in superstring at one-loop level (torus amplitude) is actually vanishing. Indeed, this agrees with the conjecture in \cite{SU} as speculated from the absence of the renormalization of Newton constant. We will leave a direct confirmation of this expectation as an important future problem.

\section{Conclusions}

The main purpose of this paper was to study the conical entropy $S_A$ in string theory. The conical entropy is defined as the entanglement entropy computed by using the replica method and it is not guaranteed to be positive as it includes the surface term contributions.
In terms of black hole entropy for the Rindler horizon, our conical entropy from string theory one-loop amplitudes corresponds to the leading quantum corrections \cite{SU}.

As a warmup, we started with a computation of the conical entropy for free fields with any higher spins. By using an interpolation between of the replicated space and the orbifold $C/\mathbb{Z}_{N}$, we find a new simple way to evaluate the conical entropy for free higher spin fields. Not only our method successfully reproduces the know results \cite{Kab,FM} for spin $0,1/2,1,3/2$ and $2$, but also it offers a simple formula for any higher spins.

Next we calculated the conical entropy for open string theory. The evaluation is easier than that for closed string as we only need to perform a single summation corresponding to the
$\mathbb{Z}_{N}$ projection. Though we manage to find an explicit expression in terms of summations over massive modes, the conical entropy turns out to be divergent even in superstring unless some miraculous cancellations occur. This is due to the UV divergence and can be regarded as the summation of area law divergences over all open string modes. The fact that the conical entropy is divergent implies that this calculation is not trustable. To resolve this issue, we will need to take into account the back reactions of open strings to the closed string sector as we do so in the AdS/CFT setup. We will leave this for a future work.

Finally we studied the conical entropy for type II closed string. We focused on the contributions
from the one-loop torus vacuum amplitudes, which are interpreted as the entanglement entropy of free superstring theory based on the replica method. One of the most interesting aspects of this calculation in closed superstring is that the moduli integral is limited to the fundamental region as opposed to the open superstring. This is the basic mechanism of the UV cut off in string theory and can be a possibility to get a finite conical entropy.
However, a technical problem arises immediately in order to evaluate the conical entropy explicitly. This is because we need to perform double sum at the same time: one is the orbifold projection and the other is the summation over twisted sectors.

To remedy this problem, we turned to a modified quantity called the twisted conical entropy $\ti{S}_A$, which is defined as a conical entropy not for the orbifold $C/\mathbb{Z}_{N}\times R^8$ but for the Melvin orbifold $(C\times S^1)/\mathbb{Z}_{N}\times R^7$. In this case we can first rewrite the double sum into a single sum by employing the folding trick, which changes the integral on the fundamental region into that on the strip removing one of the two summations. Owing to this trick we can evaluate the twisted conical entropy $\ti{S}_A$ and can confirm that this quantity is UV finite as expected in superstring theory. Note also that our twisted (Renyi) conical entropy $\ti{S}^{(n)}_A$ is well-defined and finite for any $n=1/N$ only when the radius $R$ is large enough compared with the string scale $\s{\al}$ so that there is no closed string tachyons. On the other hand, if $R$ is very small, we find $\ti{S}^{(n)}_A$ gets divergent due to the tachyons. However we can still define the $N=1$ limit $\ti{S}_A$ as a finite quantity even in this case. Since the equations of motion in string theory are fully satisfied at $N=1$ including the tachyon fields, we believe that the effect of closed string tachyon condensation appears in the partition function at order $O((N-1)^2)$ and thus do not affect $\ti{S}_A$.
Moreover we observed that the analytical continuation of the partition function does not have any UV divergences for $N<1$.

In order to go back to our original problem of computing the conical entropy, we need to take the small radius limit of Melvin background. We expect that the leading contribution of twisted conical entropy $\ti{S}_A$ in this limit is reduced to the conical entropy $S_A$. We find that in this limit, $\ti{S}_A$ doed not have any contributions which are proportional to the area (or any higher power) of the boundary $\de A$. Therefore this result indirectly seems to support the conjecture \cite{SU} that (quantum corrections to) the conical entropy $S_A$ in type II closed superstring is vanishing. It will be a very interesting future problem to discover a direct method to evaluate the conical entropy in superstring.

\section*{Acknowledgements}

We would like to thank Stuart Dowker, Mukund Rangamani, Kathrin Bringmann, Victor Kowalenko, Juan Maldacena, Tatsuma Nishioka, Masahiro Nozaki and Noburo Shiba for useful
discussions. SH is supported by JSPS postdoctoral fellowship for foreign researchers and by the National Natural Science Foundation of China (No.11305235). TN is supported by JSPS fellowship. TT is supported by JSPS Grant-in-Aid for Scientific Research (B) No.25287058 and by JSPS Grant-in-Aid for Challenging
Exploratory Research No.24654057. TT is also supported by World Premier International Research Center Initiative (WPI Initiative) from the Japan Ministry
of Education, Culture, Sports, Science and Technology (MEXT).

\appendix

\begin{appendix}

\section{Folding Trick}

We are interested in the following generic formula
\ba
Z(A,B)&=&\int_{F} \f{(d\tau)^2}{(\tau_2)^5}
\sum_{n,m\in {\mathbb Z}}\
\f{|\theta_{1}({B(n-m \tau)\over 2}|\tau)|^8}{|\eta(\tau)|^{18}
|\theta_{1}({B(n-m \tau)}|\tau)|^2}\times \exp\left[-{A|n-m\tau|^2\over \tau_2}\right]. \label{identity}
\ea
Where $F$ stands for the fundamental domain in complex plane and $A, B$ are arbitrary positive constants.
We will do the modular transformation as
\ba \tau\rightarrow \tau'={a\tau+b\over c\tau+d}
\ \ \ \ (ad-bc=1)
\ea for (\ref{identity}).
We can obtain following transformation rules:
\ba \tau'_2&=& {\tau_2\over |c\tau+d|^2} \label{transformationrule1}\\
n-m \tau'&=&{(n c-ma)\tau+(n d-m b)\over |c\tau+d|^2} \label{transformationrule2}\\
|\eta(\tau')|&=& |(c\tau+b)|^{1\over 2}|\eta(\tau)|\label{transformationrule3}\\
\theta_{1}\left({B(n-m \tau')}|\tau'\right)&=& \theta_{1}\left({B (n c-m a )\tau+(n d-m b)\over c\tau+d}|{a\tau+b\over c\tau+d}\right)\label{transformationrule4}\\
\theta_{1}({B r \over c \tau+d}|{a\tau+b\over c\tau+d})&=& \xi_r(c\tau+d)^{{1\over 2}}e^{i \pi {(B r)^2 c\over c\tau+d}}\theta_{1}(B r|\tau)\label{transformationrule5}.
\ea
Where $\xi_r^8=1$ and $r$ are arbitrary positive integers and (\ref{transformationrule5}) are related to the nice property of modular function $\theta(\mu|\tau)$.
We can make use of these transformation rules to the integrand given in (\ref{transformationrule1})(\ref{transformationrule2})
(\ref{transformationrule3})(\ref{transformationrule4}):
\ba
Z(A,B)&=&\int_{F} \f{(d\tau)^2}{(\tau_2)^5}
\sum_{n,m\in {\mathbb Z}}\
\f{|(c\tau+d)|^6|\theta_{1}({B (n c-m a )\tau+(n d-m b)\over 2(c\tau+d)}|{a\tau+b\over c\tau+d})|^8}{|(c\tau+d)|^{9}|\eta(\tau)|^{18}
|\theta_{1}({B (n c-m a )\tau+(n d-m b)\over c\tau+d}|{a\tau+b\over c\tau+d})|^2}\times\no &  & \exp\left[-{A| (n c-m a )\tau+(n d-m b)|^2\over \tau_2}\right]. \no \label{Pidentity}
\ea
We need to sum over all $n,m\in {\mathbb Z}$ and choose the integers $(a,b,c,d)$ such that $n c-m a=0$. Since $a$ and $c$ are coprime, we can express $n$ and $m$ as $n=ra$ and $m=rc$, where $r=[m,n]$ is the greatest common divisor of $m$ and $n$.
Here we just replace the double sum in the following way
\ba \sum_{m,n}[...]=\sum_{r}\sum_{[m,n]=r}[...],\ea
 following papers \cite{Polchinski:1985zf}\cite{O'Brien:1987pn}\cite{McClain:1986id}.

 Therefore, we can simplify the integration as
\ba
Z(A,B)\!&=&\!\int_{F} \f{(d\tau)^2}{(\tau_2)^5}
\sum_{r\in {\mathbb Z}}\sum_{[m,n]=r}\
\f{|\theta_{1}({B r \over 2(c\tau+d)}|{a\tau+b\over c\tau+d})|^8}{|\eta(\tau)|^{18}
|\theta_{1}({B r\over c\tau+d}|{a\tau+b\over c\tau+d})|^2}  \exp\left[-{A r^2\over \tau_2}\right]{1\over |c\tau+d|^3},\label{Pidentity1}\\
&=&\!\int_{S} \f{(d\tau)^2}{(\tau_2)^5}
\sum_{r\in {\mathbb Z}}\
\f{|\theta_{1}({B r \over 2}|{\tau})|^8}{|\eta(\tau)|^{18}
|\theta_{1}({B r}|{\tau})|^2}\times  \exp\left[-{A r^2\over \tau_2}\right].\label{Pidentity2}
\ea
Where we have made use of $(n d-m b)=r$ in (\ref{Pidentity1}) and applied transformation rule (\ref{transformationrule5}) in the final identity (\ref{Pidentity2}). The final step in (\ref{Pidentity2}) is explained by applying the arguments in \cite{Polchinski:1985zf}\cite{O'Brien:1987pn}\cite{McClain:1986id}  and the modular invariance of (\ref{identity}). The sum over all pairs $(m,n)$ with fixed $r=[m,n]$ enlarges \ the fundamental domain $F$ into the strip domain $S$ defined by $-{1\over 2}<\tau_1<{1\over 2}, \tau_2>0$. In this way, we can simplify the double summation in (\ref{identity}) as a single summation finally.

\section{Summation}
In this appendix, we would like to show some details about how to do the summation appeared in this paper. Especially here we want to explain the summation (\ref{pos}) in section (5.2) as an example. This kind of procedure can also be applied to other summations appeared in this paper and we do not repeat these details here.
The partition function has following form
\ba
\int_{S} \f{(d\tau)^2}{(\tau_2)^5}
\sum_{\alpha\in {\mathbb Z}}\sum_{\beta=0}^{N-1}\
{\sin^8({\beta \pi\over N})\over \sin^2({2\beta\pi\over N })}\times  \sqrt{\tau_2}\exp\left[-{\pi^2 \tau_2 \alpha^2\over A N^2}+{2\pi i \beta \alpha\over N}\right]\label{Pidentityy3}\\
 \ea where we made use of Possion resummation formula
 \be
 \sum_{\alpha}{1\over \sqrt{\tau_2}} \exp\Big(-{\pi A(N\alpha^2+\beta)^2\over \pi \tau_2}\Big)=\sum_{\alpha}\exp\Big(-{\pi^2 \tau_2 \alpha^2\over AN^2}+{2\pi i \beta \alpha\over N}\Big).
 \ee
In this part, our calculation will be reduced to the following summation
\ba S_{ker}&= &\sum_{\alpha=-\infty}^{\infty}\sum_{\beta=0}^{N-1}{\sin^8({\beta \pi\over N})\over \sin^2({2\beta\pi\over N })}\exp({2\pi i\beta \alpha\over N})e^{-{\pi \tau_2 \alpha^2\over A N^2}}\no
&=&\sum_{\alpha=-\infty}^{\infty}\sum_{\beta=0}^{N-1}\Big(-\frac{1}{64} \cos \left(\frac{2 \pi  \beta (\alpha -2)}{N}\right)-\frac{15}{32}  \cos \left(\frac{2 \pi  \alpha  \beta }{N}\right)+\frac{1}{8} \cos \left(\frac{2 \pi  \beta (\alpha -1)}{N}\right)\nonumber\\
&+&\frac{1}{8} \cos \left(\frac{2 \pi  \beta (\alpha +1)}{N}\right)-\frac{1}{64} \cos \left(\frac{2 \pi  \beta (\alpha +2)}{N}\right)+{1\over 4}\cos({2\pi \alpha\beta\over N})\sec^2({\pi\beta\over N})\Big)e^{-{\pi \tau_2 \alpha^2\over A N^2}}\no
&=& \sum_{\alpha=1}^{\infty}{1\over 2}S_3(N,\alpha \mod N, 1)e^{-{\pi \tau_2 \alpha^2\over A N^2}}+{1\over 4}N^2-{15N\over 32}\no
&&+ \sum_{\alpha=1}^{\infty}\Big(-\frac{15}{32}  \sin (2 \pi  \alpha ) \cot \left(\frac{\pi  \alpha }{N}\right)\Big)e^{-{\pi \tau_2 \alpha^2\over A N^2}}.\no\label{onesum}\ea Where we have employed the formula $\sum_{\beta=0}^{N-1} \cos (\beta x)={1\over 2}\Big(1+{\sin(N-{1\over 2})x \over \sin ({x\over 2})}\Big)$

and $\vartheta _3(u,q)=2 \sum _{n=1}^{\infty } q^{n^2} (\cos  (2 n u))+1$ as well as the definition of $S_3(N,\alpha \mod N, 1)$ given by \cite{summation} as follows
\ba S_{3}(q,r,n):&=& \sum_{p=1}^{q-1} \cos({2 r p\pi \over q})\csc^{2n}\left({p\pi\over q}\right)={(-1)^{n-1}2^{2n}\over 2n!}\no
&&\sum_{\alpha=0}^{n} \left(
\begin{array}{c}
 2n \\
 2 \alpha \\
\end{array}
\right)B_{2\alpha}\left({r\over q}\right) B_{2n-2\alpha}^{(2n)}(n)q^{2\alpha}. \ea
Here we just focus on the left summation in (\ref{onesum}),
\ba \sum_{\alpha=1}^{\infty}{1\over 2}S_3(N,\alpha\mod N, 1)&=&\sum_{\alpha=1}^{\infty}\sum_{k=0}^{1}\left(\left(
\begin{array}{c}
 2 \\
 2 k \\
\end{array}
\right) B_{2k}({\{{\alpha\over N}\}})B_{2k}^{(2)}(1)N^{2k}\right)e^{-{\pi \tau_2 \alpha^2\over A N^2}}\no
&=&\sum_{\alpha=0}^{\infty}\left[-{1\over 6}B_0\left(\{{\alpha\over N}\}\right)+N^2 B_2(\{{\alpha\over N}\})\right]e^{-{\pi \tau_2 \alpha^2\over A N^2 }}.
\label{summationofS}\ea  Where $\{x\}$ denotes the fractional part of real number $x$ and the formal definition is $\{x\}=x-\lfloor x\rfloor, x\in \mathbf{R}$. The floor function $\lfloor x\rfloor$ gives the largest integer
less than or equal to $x$.
The Bernoulli polynomials $B_{n}^{m}(x)$ of order $m$ and of degree $n$ in $x$ (sometimes also called the higher order or generalized Bernoulli polynomials) can be defined
by following relation
\ba
\left({t\over e^t-1}\right)^m e^{x t}=\sum_{n=0}^{\infty}B_{n}^{(m)}(x){t^n\over n!} \text{ }\text{ }(|t|<2\pi; m \in \mathbf{N}_0:=N\bigcup \{0\}).
\ea
In the last step of (\ref{summationofS}) , we have used fact $B_{2}^{(2)}(1)=-{1\over 6},\  B_{0}^{(2)}=1$. The Bernoulli numbers $B_{n}^{(m)}(x)$ of order $m$ and degree $n$  defined by
\ba
B_{n}^{(m)}(x)=\sum_{k}^n B_{k}^{(m)}x^{n-k}
\ea with $B_{n}^{(m)}=B_{n}^{(m)}(0)$. Combined (\ref{onesum}) and (\ref{summationofS}),
the partition function of string theory on the Melvin background $Z_{closed}\left[(C\times S^1)/\mathbb{Z}_N\times R^7\right]$ defined in (\ref{melmel}) can be evaluated as follows in large $\tau_2$ limit:
\ba
&& Z_{closed}\left[(C\times S^1)/\mathbb{Z}_N \times R^7\right] \no
&&\sim\int_{S} \f{(d\tau)^2}{(\tau_2)^{9\over 2}}\Big({1\over 4}N^2-{15N\over 32}+ \sum_{\alpha=1}^{\infty}\Big(-{1\over 6}+{N^2\over 6}(1-6\{{\alpha\over N}\}
+6\{{\alpha\over N}\}^2)\no
&& \ \ \ \ \ \ \ \ \ \ \ \
-\frac{15}{32}  \sin (2 \pi  \alpha ) \cot \left(\frac{\pi  \alpha }{N}\right)\Big)e^{-{\alpha'\tau_2 \alpha^2\over R^2 N^2}}\Big).\ea
Where we used $A={\pi R^2\over \alpha'}$ in terms of our convention. Therefore, the leading contribution is $\int_{S} \f{(d\tau)^2}{(\tau_2)^{9\over 2}}\Big({1\over 4}N^2-{15N\over 32}\Big)$ which is consistent with (\ref{gn}) up to leading order $O(\exp({- \tau_2}))$.

\end{appendix}

%\newpage
%%%%%%%%%%%%%%%%%%%%%%%%%%%%%%%%%%%%%%%
%%%%%%%%%%%%%%%%%%%%%%%%%%%%%%%%%%%%%%%

\end{document}